\documentclass{elsart}

 \usepackage{graphics}
 \usepackage{graphicx}
\usepackage{amssymb}
\usepackage{bm}

\begin{document}




\title{Measurement of the neutron electric dipole moment by crystal diffraction}

\author[PNPI]{V.~V. Fedorov}, \author[ILL]{M. Jentschel}, \author[PNPI]{I.~A. Kuznetsov}, \author[PNPI]{E.~G. Lapin}, \author[ILL]{E. Lelievre-Berna}, \author[ILL]{V. Nesvizhevsky}, \author[ILL]{A. Petoukhov}, \\ \author[PNPI]{S.~Yu. Semenikhin}, \author[ILL]{T. Soldner}, \author[ILL]{F. Tasset},\\ \author[PNPI]{V.~V. Voronin}, \author[PNPI]{Yu.~P. Braginetz}

\address[PNPI]{Petersburg Nuclear Physics Institute, Gatchina,
St.~Petersburg, Russia}
\address[ILL]{Institut Laue-Langevin, Grenoble, France}

{\footnotesize Email: vvv@pnpi.spb.ru (Voronin V.V.)}

\begin{abstract}
An experiment using a prototype setup to search for the neutron electric
dipole moment by measuring spin-rotation in a non-centrosymmetric crystal
(quartz) was carried out to investigate statistical sensitivity and systematic
effects of the method. It has been demonstrated that the concept of the
method works. The preliminary result of the experiment is
$d_{\rm n}=(2.5\pm 6.5)\cdot 10^{-24}$~e$\cdot $cm. The experiment showed
that an accuracy of  $\sim 2.5\cdot 10^{-26}$~e$\cdot $cm can be obtained
in 100 days data taking, using available quartz crystals and neutron
beams.
\end{abstract}

\section{Introduction}
\label{Intr}
The electric dipole moment of the neutron (nEDM) is a very sensitive probe
for CP violation beyond the Standard Model of particle physics
\cite{Ritz,RamseyMusolf}. The most precise experiments today use Ramsey's
resonance method and ultra-cold neutrons (UCNs) \cite{pnpiedm,edmlast}.
Further progress is presently limited by systematics \cite{pendlebury2004}
and the low density of UCNs available. Here we discuss an alternative
approach based on spin rotation in non-centrosymmetric crystals.

The statistical sensitivity of any experiment to measure the nEDM is
determined by the product $E\tau \sqrt{N}$,
where $\tau$ is the duration of the neutron interaction with the electric
field $E$ and $N$ the number of the counted neutrons.
New projects to measure the nEDM with UCNs aim to increase the UCN density
and thus $N$ by orders of magnitude (see \cite{golub2005} for a recent
overview). In contrast, experiments with crystals exploit the electric
field inside matter, which for some crystals can be by a few orders of
magnitude higher than the electric field achievable in vacuum.

EDM experiments with absorbing crystals were pioneered by Shull and Nathans
\cite{shull1967}. Their experiment was based on the
interference of the electromagnetic amplitude with the imaginary part of the
nuclear one. Abov with his colleagues \cite{Abov1966} were the first who paid
attention to the presence of a spin dependent term due to the interference of
nuclear and spin-orbit parts of the scattering amplitude in the interaction of
neutrons with a non-centrosymmetric non-absorptive crystal. Spin-rotation in
non-centrosymmetric crystals due to such
interference effects as a way to search for a nEDM was first discussed by
Forte \cite{forte1983}. The corresponding spin-rotation effect due to
spin-orbit interaction was experimentally tested by Forte and Zeyen
\cite{ForteZeyen}. The authors of \cite{grav,dfield} have shown that the
interference of the nuclear and the electromagnetic parts of the
scattering amplitude leads to a constant strong electric field,
acting on a neutron during all time of its movement in the
non-centrosymmetric crystal. This field was measured in a Laue geometry
diffraction experiment \cite{dfield}, in agreement with the calculated value. 

The spin rotation can be measured in Bragg
\cite{forte1983,ForteZeyen,Baredm} and Laue
\cite{dedm,polart,PhysB2001,dedm1} diffraction geometry. In this paper, we compare
both geometries and present preliminary results of a test experiment
in Bragg geometry. We show that the sensitivity of an optimized experiment
in Bragg geometry can compete with the most sensitive published UCN nEDM
measurements.


\section{Comparison of Laue and Bragg diffraction geometry}

A detailed recent study of a nEDM measurement in Laue geometry can be found in
\cite{LDM_sens}. The main advantage of this scheme is the possibility to
increase the time $\tau$ of neutron passage through the crystal using
Bragg angles $\theta_{\rm{B}}$ close to $\pi/2$ \cite{dedm}. In this way,
times close to the time of neutron absorption $\tau_{\rm{a}}\approx 1$ ms 
were obtained in short crystals \cite{PhysB2001,tfjetpl}.
However, in order to suppress systematic effects due to the Schwinger
interaction (spin-orbit coupling), the method relies on an effective
depolarization of the
neutron beam by Schwinger-rotating the two components of the neutron wave by
$\pm\pi/2$. This fixes the thickness of the crystal to \cite{dedm1}:
\begin{equation}\label{L0}
L_0=\frac{\pi m_{\rm{p}}c^2}{2\mu_{\rm n}eE_{\bm g}}.
\end{equation}
$E_{\bm g}$ is the electric field affecting the neutron for the exact Bragg
condition for the crystallographic plane $\bm g$ ($\bm g$ is a reciprocal
lattice vector)
and $m_{\rm{p}}$ the proton mass \cite{dfield}. Therefore the sensitivity
of the method cannot be increased by using crystals with a higher electric
field $E_{\bm g}$ since the crystal length and thus the time $\tau$ would need
to be reduced. On the other hand, it is impossible to increase the sensitivity
using Bragg angles extremely close to $\pi/2$, because of a
resulting decrease of the neutron count rate for such angles \cite{LDM_sens}.
In the experiment \cite{LDM_sens}, also a strong sensitivity of the
method to crystal deformations was found.

The main advantage of the Bragg diffraction scheme \cite{forte1983,Baredm}
is that the electric field acting on the neutron depends on the deviation
of the neutron trajectory from the Bragg condition. This allows us to
control value and even sign of the electric field and
makes new tests of systematic effects possible. On the other hand,
the time $\tau$ that the neutron spends in the crystal cannot be
increased by using Bragg angles close to $\pi/2$ as it depends on the total
neutron velocity $v$ and not on the velocity component parallel
to the crystallographic planes as in the Laue case. However, this
disadvantage can be obviated by increasing the crystal thickness, in principle.

A first experiment in Bragg geometry was reported in \cite{ForteZeyen},
measuring the neutron spin rotation due to Schwinger interaction in quartz.
The experimental value of the spin rotation angle was a few times
less than the theoretical expectation \cite{ForteZeyen}. The main
experimental difficulty was to obtain monoenergetic neutrons with a
well-defined small deviation from the Bragg condition. This was
solved by placing the monochromator in a strong magnetic field, but
this field may cause systematic errors.

Here we propose and use a very simple solution of the problem to obtain these
monoenergetic neutrons.

\section{Two-crystal scheme}

Let's consider the symmetric Bragg diffraction case. A neutron falls on the
crystal in a direction close to the Bragg one for the crystallographic
plane $\bm g$. The deviation from the exact Bragg condition is described
by the parameter
$\Delta E_{\bm g}=E_{\bm k}-E_{{\bm k}_{\bm g}}$, where $E_{\bm k} = \hbar^2k^2/2m$
and $E_{{\bm k}_{\bm g}}=\hbar^2|{\bm k} + {\bm g}|^2/2m$ are the energies
of a neutron in the states $\vert {\bm k} \rangle $ and
$\vert {\bm k}+{\bm g}\rangle$, respectively.

In this case the neutron wave function inside the crystal in first order
of perturbation theory can be written \cite{Sch91}
\begin{equation}\label{Psi}
\psi ({\bm r}) = e^{ - i\;{\bm k\bm r}} + a \cdot e^{-i({\bm k} + {\bm g}){\bm r}},
\end{equation}
where
\begin{equation}\label{a}
a = \frac{\left| {V_{\bm g} } \right|}{E_{\bm k} - E_{{\bm k}_{\bm g} }} = \frac{\left| {V_{\bm g} } \right|}{\Delta E_{\bm g}}.
\end{equation}
Here $V_{\bm g} $ is the ${\bm g}$-harmonic of the neutron-crystal interaction potential.
For simplicity we consider the case $a \ll 1$, permitting to use perturbation theory.

\begin{figure}[htbp]
    \centering
        \includegraphics[width=0.8\textwidth]{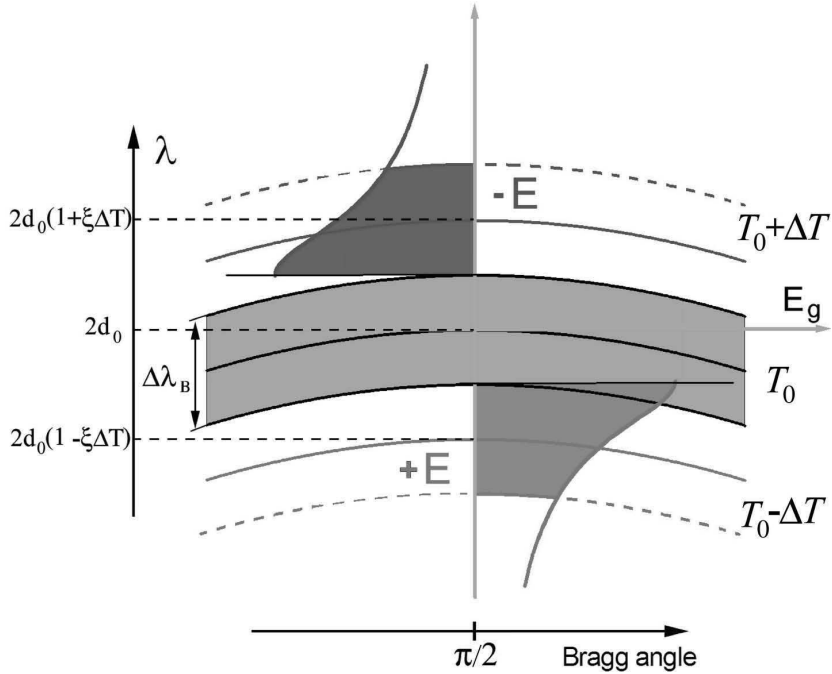}
    \caption{Electric field for neutrons close to the Bragg condition 
      (Bragg angle $\pi/2$ and $\lambda = 2d_0$) in
      a non-centrosymmetric crystal: The electric field has a pole for
      neutrons exactly fulfilling the Bragg condition. By choosing the
      deviation from the Bragg condition, value and sign of the electric
      field can be selected. We use Bragg reflection in a second crystal
      with a different temperature $T_0\pm\Delta T$ to select neutrons
      with a certain deviation.}
    \label{fig:2}
\end{figure}

The electric field acting on the neutron in the crystal is \cite{Sch91}
\begin{equation}\label{E}
{\bm E} = {\bm E}_{\bm g} \cdot a, \label{Eq:E}
\end{equation}
where ${\bm E}_{\bm g}$ is the interplanar electric field for the exact Bragg condition.
A nonzero nEDM $d_{\rm n}$ results in neutron spin rotation by the angle
\begin{equation}
\varphi _{\rm EDM} = \frac{2 E \cdot d_{\rm n} \cdot L}{\hbar v_\bot},
\label{Eq:fid}
\end{equation}
where $L$ is the length of the crystal and $v_ \bot$ the component of
the neutron velocity perpendicular to the crystallographic plane.
Obviously, sign and value of the electric field (\ref{E}) are determined  by
sign and value of the deviation  from the exact Bragg condition $\Delta E_{\bm g}$. This is illustrated in Fig.~\ref{fig:2}.

The presence of the electric field will lead to the appearance of a Schwinger magnetic field 
\begin{equation}\label{Hgs}
  {\bm H}_{\rm S}= [{\bm E}\times {\bm v}]/c,
\end{equation}
where $\bm v$ is the neutron velocity. The corresponding spin rotation
angle is \cite{PhysB2001}
\begin{equation}
\varphi_{\rm S} = \frac{2\mu H_{\rm S} L} {\hbar v_\bot}.
\end{equation}

Note that the Schwinger effect disappears for Bragg angles of $\pi/2$
as ${\bm E} \parallel {\bm v}$ in this case:
\begin{equation}
\varphi_{\rm S} = \frac{2 E \mu L v_\parallel}{c\hbar v_\bot} =
  \frac{2 E \mu L}{c\hbar}\cot\theta_{\rm B} \mathrel{\mathop{\kern0pt\longrightarrow}\limits_{\theta _{\rm B} \to \pi / 2}} 0,
\label{Eq:fis}
\end{equation}
where  $v_\parallel$ is the component of the neutron velocity parallel to the crystallographic plane.
This can be used in the nEDM experiment to suppress systematic effects due to
Schwinger interaction.

In the experiment \cite{JETPLetNoptic}  the effect of neutron spin rotation in neutron optics for large ($ \sim 10^3 - 10^4$ Bragg
widths) deviations from the exact Bragg condition have been observed. The measured effect has coincided with the theoretical expectation.

\begin{figure}[htbp]
    \centering
        \includegraphics[width=0.8\textwidth]{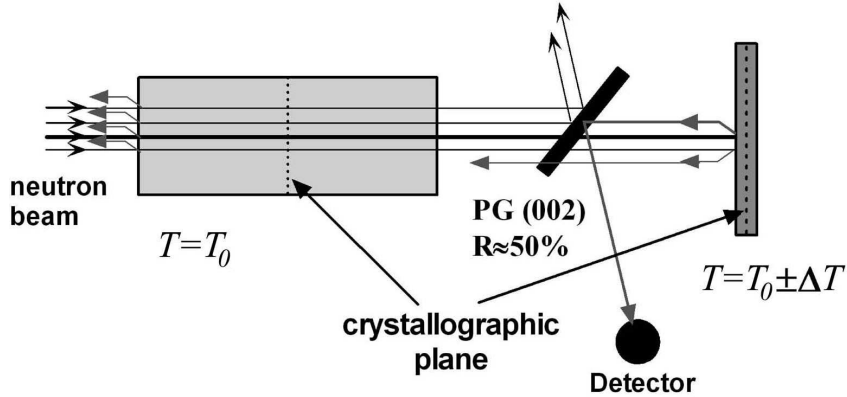}
    \caption{Two crystals are in parallel position. The crystal deviation
    parameter $\Delta E_g $ for the neutrons reflected by the small crystal is
    determined by the temperature difference $\Delta T$.}
    \label{fig:1}
\end{figure}

In order to tune the electric field $\bm E$, we have to select neutrons with
a well-defined deviation $\Delta E_{\bm g} $ from the Bragg condition. For this purpose,
we use two separate crystals in parallel orientation (see Fig.~\ref{fig:1}).
By heating or cooling the second (small) crystal the
interplanar distance $\Delta d$ changes and, therefore, the energy of the
reflected neutrons. The second crystal selects only the neutrons
corresponding to its own Bragg condition from the whole beam passing through the
first crystal. The deviation parameter and, accordingly, value and sign of the
electric field having affected these neutrons in the first crystal depend
directly on the temperature difference between the two crystals (see
Fig.~\ref{fig:2}).

The value of the Bragg width $\Delta \lambda_{\rm B} $ for the (110) quartz
plane ($d=2.45$\AA) is $\Delta \lambda_{\rm B} / \lambda \approx 10^{ - 5}$.
To shift the reflected wavelength by one Bragg width, $\Delta d / d$ should
have the same value, corresponding to a temperature difference of
$\Delta T \approx \pm 1~{\rm K}$ (linear coefficient of thermal expansion for
quartz $\xi\equiv\Delta L / L \approx 10^{-5}~{\rm K}^{-1}$).
Note that a common variation of the two crystals'
temperatures in itself does not influence the value of the electric
field; only the temperature difference between the two crystals needs to be
controlled.

\section{Experimental setup}

\begin{figure}[htbp] 
    \centering
        \includegraphics[width=1\textwidth]{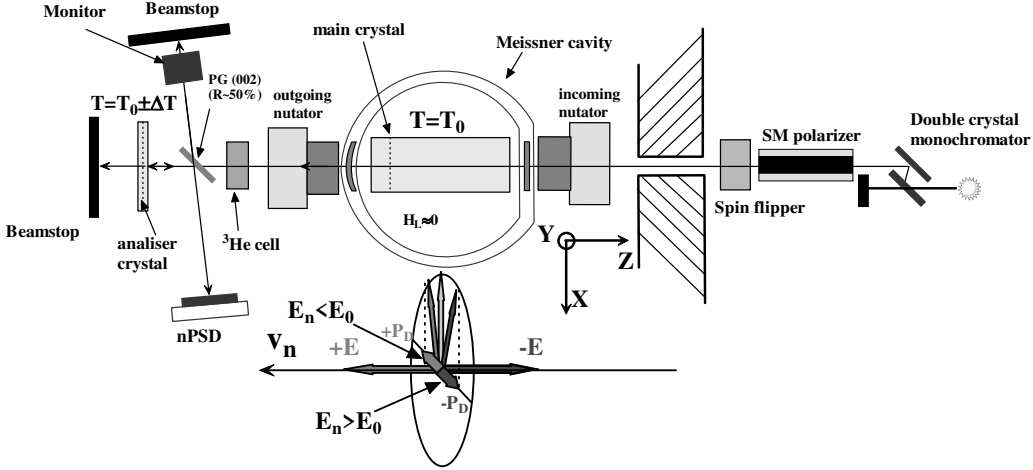}
    \caption{Scheme of the experiment. The neutrons, coming from the
      right, pass a pyrolytic graphite monochromator (adjusted to 4.91~\AA),
      a super mirror polarizer, and a resonance spin flipper.
      The incoming nutator and a coil inside the Meissner cavity orient the
      incident neutron polarization vector
      in space before the main crystal. Correspondingly,
      a second coil and the outgoing nutator
      select the component of the outgoing neutron polarization that is
      analyzed by a $^3$He spin filter. The analyzer crystal
      back-reflects neutrons according to its Bragg condition.
      The pyrolytic graphite PG (002) with about 50 \%
      reflectivity directs the reflected neutrons to the nPSD.
      It also unavoidably reflects a part of the beam after the spin filter
      which was used to monitor the beam intensity. 
      }
    \label{fig:12} 
   \end{figure}

A test experiment was carried out at the facility for particle physics
with cold neutrons PF1B \cite{abele2006} of the Institut Laue-Langevin.
The scheme of the experiment is shown in  Fig.~\ref{fig:12}.
We used the (110) reflex of quartz. The length of the main crystal was 14~cm.
As mentioned above the backscattering geometry (Bragg angle close to $\pi $/2)
allows us to suppress effectively the Schwinger effect.

To measure the nEDM we should direct the initial neutron spin along the
X axis (see Fig.~\ref{fig:12}) and investigate the dependence of the Y component
of the final polarization vector (i.e. the element $M_{XY}$
of the polarization tensor) on the temperature difference between the two
crystals.
The main systematic error of this experiment is due to the residual
Schwinger effect. It can be investigated by measuring the polarization
component along the Z axis (elements $M_{XZ}$ and $M_{YZ}$
of the polarization tensor). Therefore the measurement of a single projection
of the final polarization vector only is not sufficient. We used the
3-D polarization
analysis device CRYOPAD \cite{Cryopad} to measure the whole polarization tensor.
In first order the measured difference of the polarization tensor after 
reversing the electric field is
 \begin{equation}
\hspace{-1cm} {\bf{\Delta M}}  = g_{\rm n} \tau _0\left(
 \begin{array}{ccc}
   0 &
   -\left(H^z\frac{\Delta\tau}{\tau_0}+H_{\rm EDM}\right) &
   \left(H^y \frac{\Delta\tau}{\tau_0}+H_{\rm S}^y\right) \\
   \left(H^z \frac{\Delta\tau}{\tau_0}+H_{\rm EDM}\right) &
   0 &
   -\left(H^x\frac{\Delta\tau}{\tau_0}+H_{\rm S}^x\right)\\
   -\left(H^y\frac{\Delta\tau}{\tau_0}+H_{\rm S}^y\right) &
   \left(H^x \frac{\Delta\tau}{\tau_0}+H_{\rm S}^x\right)&
   0 
  \end{array}
\right),
\end{equation}
{\noindent where $\tau _0=(\tau _++\tau _-)/2$, $\Delta\tau =(\tau _+-\tau _-)/2$,
and $\tau _+$ and  $\tau _-$ are the times the neutrons stay in the crystal
for the positive and the negative electric field, respectively. $H^i$ are
the components of the residual magnetic field and $H_{\rm S}^i$ the
components of the Schwinger magnetic field ${\bm H}_{\rm S}$.
$g_{\rm n}=2\mu_{\rm n}/\hbar=1.8\cdot 10^4~{\rm G}^{-1}{\rm s}^{-1}$ is the neutron gyromagnetic ratio, $H_{\rm EDM}=E d_{\rm n}/\mu_{\rm n}$ the effective magnetic field
corresponding to the electric field $E$ and the nEDM $d_{\rm n}$.
For $E=1 \cdot 10^8$~V/cm and $d_{\rm n}=10^{-26}$~e$\cdot$cm,
$H_{\rm EDM}=1.7\cdot 10^{-7}$~G.}

Two main effects can simulate a nEDM: 
\begin{enumerate}
	\item Schwinger effect. It results in nonzero components
	$\Delta M_{XZ}=-\Delta M_{ZX}\approx  H_{\rm S}^y g_{\rm n} \tau _0$ and
	$\Delta M_{YZ}=-\Delta M_{ZY}\approx  H_{\rm S}^x g_{\rm n} \tau _0 $
	but does not contribute directly to the components
	$\Delta M_{XY}=-\Delta M_{YX}$ related to the nEDM.
	However, false effects can arise from imperfections of the 3-D
	polarization analysis device. These effects can be investigated
	by measuring the residual Schwinger effect via $\Delta M_{XZ}$ or
	$\Delta M_{ZX}$.
	For example, if the angular accuracy of the 3-D analysis is
	$10^{-3}$ and the Schwinger effect is suppressed down to a level of
	$10^{-4}$~G (quite realistic numbers), then the residual false effect
	will be on the level $10^{-7}$~G, corresponding to 
	$d_{\rm n}\sim 6\cdot 10^{-27}$~e$\cdot$cm.
	\item Residual magnetic field. Effects due to nonzero magnetic
	field do not compensate in the difference tensor since the neutrons
	for both signs of the deviation parameter $\Delta E_{\bm g} $ have slightly
	different energies and thus spend slightly different times in the
	residual magnetic field: $\Delta\tau/\tau_0 = -\Delta\lambda/\lambda$.
	In our case $\Delta \lambda/\lambda \approx  10^{-5}$. Therefore
	the residual magnetic field in the polarization analysis device
	needs to be small:
	\begin{equation}
	  H^z \frac{\Delta \tau }{\tau _0} \ll H_{\rm EDM}.
	\end{equation}

\end{enumerate}
 
The main idea to control the contribution of the Schwinger effect to the nEDM
matrix element $\Delta M_{XY}$ is to measure the angular, i.e. the
spatial distribution of the
polarization tensor by a position sensitive neutron detector (nPSD) because
the nEDM and Schwinger effects have different angular dependences for
Bragg angles close to $\pi/2$. The nEDM effect is constant, see (\ref{Eq:fid}),
whereas the Schwinger one is proportional to
${\rm cot}(\theta _{{\rm B}})\approx (\pi/2-\theta _{{\rm B}})$,
see (\ref{Eq:fis}) and would be visible in a spatial dependence of the
components $\Delta M_{XY}$ and $\Delta M_{YX}$.

\section{Preliminary results}

  \begin{figure}[htbp]
    \centering
        \includegraphics[width=0.8\textwidth]{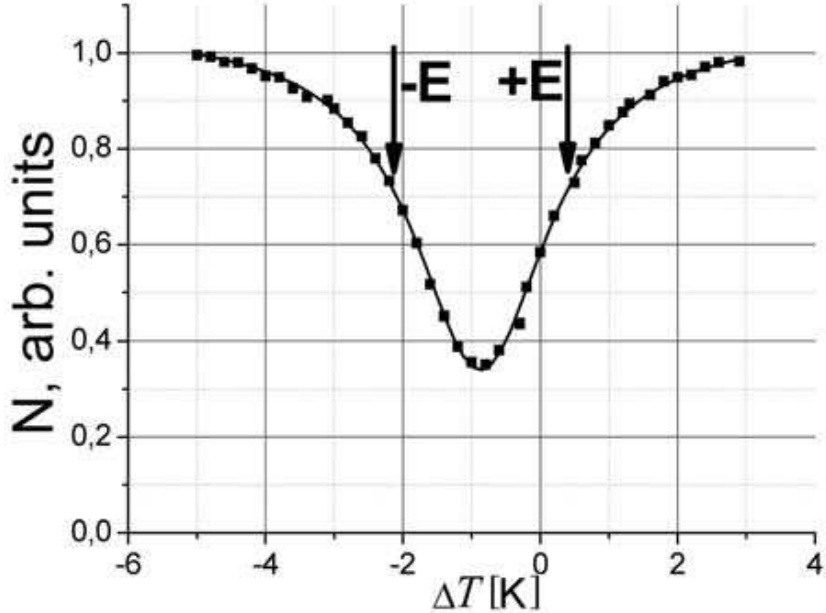}
    \caption{Dependence of registered intensity of neutrons on the
    temperature difference $\Delta T$ between the two quartz crystals.}
    \label{fig:13}
\end{figure}

The measured dependence of the integral intensity of neutrons reflected by
the second crystal on the temperature difference $\Delta T$ between the two
quartz crystals is shown in Fig.~\ref{fig:13}. The intensity
minimum corresponds to equal interplanar distances in the two crystals.
The observed difference of interplanar distances for the two crystals at $\Delta T=0$ is probably due to impurities.

In the previous measurements \cite{Bragg_test} we found the temperature
differences corresponding to maximal electric fields acting on the neutron:
$\Delta T_-=-2.0$~K and $\Delta T_+=+0.4$~K (arrows in Fig.~\ref{fig:13}). 
The corresponding electric field was determined from the slope of
the spatial dependence of $\Delta M_{ZX}$ (see \cite{panic2009} for the
graph):
\begin{equation}
 E_{\rm exp}=(0.7\pm 0.1)\cdot 10^{8}~{\rm V/cm}.
\end{equation}
This value coincides with the preliminary result obtained for the Bragg
angle $\theta_{\rm B}=86^\circ$ \cite{Bragg_test}.


The analysis of the spatial dependence of matrix element $\Delta M_{YX}$
showed no contribution from the Schwinger effect within the present
statistics. Summarizing all data, we find the value for the angle
of neutron spin rotation related to the nEDM:
\begin{equation}
\Delta \varphi_{\rm EDM}  = (0.9 \pm 2.3)\cdot 10^{- 4}.
\end{equation}
From this value, we derive the preliminary result
\begin{equation}
d_{\rm n}=(2.5\pm 6.5)\cdot 10^{-24}~{\rm e}\cdot {\rm cm.}  
\end{equation}
The error is only statistical. However, from the preliminary analysis of
systematic errors we expect that the result is limited by statistics.
The main contributions to the systematic uncertainty are due to
the Schwinger effect, the residual magnetic field inside CRYOPAD, a
small misalignment between the two crystals, and the curved exit window
of CRYOPAD. 

\section{Statistical sensitivity of a dedicated setup}

The experiment described above used existing equipment. The count rate
in the ``gray'' position of the polarization measurements was only
$\sim $ 60 n/s and the statistical
sensitivity to the nEDM was $1.6\cdot10^{-23}$~e$\cdot$cm per day. The
main limitations were:
\begin{itemize}
  \item Limited divergence acceptance. The beam divergence accepted by
    the installation was about $\pm 2.5\cdot10^{-3}$
    only. However, the beam divergence after the polarizer is about
    $\pm 1.5\cdot10^{-2}$. By installing neutron guides between the different
    elements, by shortening the installation, and by using a CRYOPAD with
    larger angular acceptance a factor of $\sqrt{18}$ in statistical
    sensitivity can be gained.
  \item Limited beam size. The used beam cross-section 5.5~cm$^{2}$ was
    limited by the size of the monochromator, the windows of CRYOPAD, and 
    the size of the crystals used. In a dedicated setup and using larger
    available
    crystals, a beam cross-section of $6\times 12~{\rm cm}^2$ is feasible,
    yielding a factor of $\sqrt{13}$ in statistical sensitivity.
  \item Limited crystal length. The interaction time of the neutrons with
    the electric field can be increased up to the absorption time constant
    in quartz, by using one large single crystal or a sequence of
    shorter crystals. We aim for a crystal length of 50~cm compared to
    14~cm used in the experiment. Taking into account the absorption,
    this will yield a factor of 2.8 in statistical sensitivity.
  \item Background, polarization analysis. Optimization of background and
    polarization analysis can yield a factor of 1.5 in statistical
    sensitivity.
\end{itemize}
With these improvements, we expect an improvement of a factor of about 65 in
statistical sensitivity, yielding a sensitivity of about
$2.5\cdot10^{-25}$~e$\cdot$cm per day. This is comparable to state-of-the-art
nEDM measurements using UCNs.

\section{Conclusions}

We have measured the nEDM by crystal diffraction in Bragg geometry.
Our preliminary result is
$d_{\rm n}=(2.5\pm 6.5)\cdot 10^{ - 24}~{\rm e}\cdot{\rm cm}$ (only statistical
error). The statistical sensitivity of the test experiment was
$1.6 \cdot 10^{-23}$~e$\cdot$cm per day. It can be improved by a factor of
65 in a dedicated setup and can reach $\sim 2.5\cdot 10^{-25}$~e$\cdot$cm per day
for available quartz crystals and neutron beam flux. The present scheme of the
experiment allows to control systematic effects related to the Schwinger
effect on-line with an accuracy better than the statistical one.
The technical requirements (e.g. accuracy of 3-D polarization analysis,
alignment of crystals) for a systematic uncertainty on the level of
$10^{-26}$~e$\cdot$cm seem achievable \cite{federovtobe}.

~

The authors would like to thank the personnel of the ILL reactor (Grenoble, France) for the technical assistance in the experiment. This work was supported by RFBR (grants No 06-02-16378-a, 09-02-00446-a).

\end{document}